\documentclass[10pt,preprint]{aastex}
\usepackage{amsmath}

\newcommand{\ncl}{\left( \frac{n_{cl}}{\mathrm{cm}^{-3}} \right)}
\newcommand{\denratio}{\left( \frac{\chi}{10^3} \right)}
\newcommand{\rcl}{\left( \frac{R_{cl}}{100\mathrm{ pc}} \right)}
\newcommand{\vshb}{\left( \frac{v_{sh,b}}{10^3\mathrm{ km/s}} \right)}

\begin{document}
%------------------------------------------------------------------------------

%\title{Hydrodynamic Simulations of Radiative Shock-Cloud Interactions}
\title{Radiative Shock-Induced Collapse of Intergalactic Clouds}

\author{P. Chris Fragile, Stephen D. Murray, Peter Anninos}
\affil{University of California,
Lawrence Livermore National Laboratory, Livermore, CA 94550}

\and

\author{Wil van Breugel}
\affil{Institute of Geophysics and Planetary Physics, Lawrence Livermore 
National Laboratory, Livermore, 94550}

%\date{{\small    \today}}
%\date{{\small   \LaTeX-ed \today}}
%-----------------------------------------------------------------------------

\begin{abstract}
Accumulating observational evidence for a number of radio galaxies
suggests an association between their jets and regions of active star
formation.  The standard picture is that shocks generated 
by the jet propagate through an inhomogeneous medium and trigger the 
collapse of overdense clouds,  which then become 
active star-forming regions.  
In this contribution, we report on recent 
hydrodynamic simulations of radiative 
shock-cloud interactions using two different cooling models:
an equilibrium cooling-curve model assuming solar metallicities 
and a non-equilibrium chemistry model appropriate for primordial gas 
clouds.
We consider a range of initial cloud
densities and shock speeds in order to quantify the 
role of cooling in the evolution.  Our results indicate that for
moderate cloud densities ($\gtrsim 1$ cm$^{-3}$) and shock Mach numbers
($\lesssim 20$), cooling processes can be highly efficient and result
in more than 50\% of the initial cloud mass cooling to below 100 K.
We also use our results to estimate the final H$_2$ mass fraction
for the simulations that use the non-equilibrium chemistry package.  This is
an important measurement, since H$_2$ is the dominant coolant for a
primordial gas cloud.  We find peak H$_2$ mass fractions of $\gtrsim
10^{-2}$ and total H$_2$ mass fractions of $\gtrsim 10^{-5}$ for the
cloud gas, consistent with cosmological simulations of first star formation.
Finally, we compare our results with the observations of jet-induced star
formation in ``Minkowski's Object,'' a small irregular starburst system
associated with a radio jet in the nearby cluster of galaxies Abell
194. We conclude that its morphology, star formation rate
($\sim 0.3M_\odot$ yr$^{-1}$) and stellar mass 
($\sim 1.2 \times 10^7 M_\odot$) 
can be explained by the interaction of a
$\sim 9 \times 10^4$ km s$^{-1}$ jet with an ensemble of 
moderately dense ($\sim 10$ cm$^{-3}$), warm ($10^4$ K)
intergalactic clouds in the vicinity of its associated radio galaxy 
at the center of the galaxy cluster.

\end{abstract}


\section{Introduction}
\label{sec:intro}

Combined optical and radio observations of a number of extragalactic
radio jets in both low-luminosity (FRI) and high-luminosity (FRII) 
galaxies have revealed interesting
correlations between the jets and apparent regions of active star
formation.  One of the first objects demonstrated to show such a
correlation was the nearest radio galaxy, Centaurus A 
\citep[e.g.][]{Blanco75}.  Other examples have been found
as the sensitivity and spatial resolution of radio and optical
telescopes has improved.  In the case of Centaurus
A, recent observations with the Hubble Space Telescope, when compared to
radio images obtained with the Very Large Array, have confirmed that there
are about half a dozen young ($< 15$ Myr) OB associations near filaments
of ionized gas located between the radio jet and a large \ion{H}{1}
cloud \citep{Mould00,rej02}. Another nearby example is ``Minkowski's Object,''
a peculiar small starburst system at the end of a radio jet emanating
from the elliptical galaxy NGC 541, located near the center of the
cluster of
galaxies Abell~194 \citep{vB85}. Star forming regions associated with
radio sources have also been found in cooling flow clusters 
\citep{McNamara02}.

Correlations between radio and optical emissions have also been observed
in the so-called ``alignment effect'' in distant ($z>0.6$) radio galaxies
\citep{Chambers87,McCarthy87}. The best studied example here is 4C41.17
at $z=3.8$, where deep spectroscopic observations have shown that the
bright, spatially extended, rest-frame UV continuum emission aligned
with the radio axis of this galaxy is unpolarized and shows P Cygni-like
features similar to those seen in star-forming galaxies \citep{Dey97}.

Collectively, these observations are most compellingly explained by
models in which shocks generated by the radio jet propagate through an
inhomogeneous medium and trigger gravitational collapse in relatively
overdense regions \citep{Begelman89,DeYoung89,Rees89}. A detailed analysis
of the jet-induced star formation in 4C41.17 has been presented by
\citet{Bicknell00}. In that object, Hubble Space Telescope images showed
a bimodal optical continuum structure parallel to the radio jet 
\citep{vB99}, strongly supporting the idea that the star formation
was triggered by sideways shocks.

Although the analytic arguments in favor of this model are compelling, 
the complex nature of this problem, including 
nonlinear effects from higher-order coupling between hydrodynamics, 
cooling, and gravity, suggest that numerical studies may provide a more
complete and detailed picture.
Furthermore, numerical studies allow us 
to probe environments that may be observationally out of reach.  
For example, such studies may be able to parameterize the conditions of 
shock-induced star 
formation in the early universe, where direct observations of 
star-forming regions are not possible.  This could be important 
to understanding the role of jets in the feedback of active galactic 
nuclei (AGN) on their environment.  Numerical simulations also give us 
insight into the role of shock-induced star formation in other 
environments such as supernova shocks and cloud-cloud collisions.

The interaction of a strong shock with a single, non-radiative cloud 
has been the subject of many numerical studies.  A thorough analysis of 
this problem and a review of relevant literature is provided by 
\citet{Klein94}.  A more recent study, considering the interaction of a 
strong shock with a system of clouds, is presented by \citet{Poludnenko02}.  
For a non-radiative cloud the passing shock ultimately destroys 
the cloud within a few dynamical timescales.  Destruction 
results primarily from 
hydrodynamic instabilities at the interface between the cloud and 
the post-shock background gas.

Less well studied is the case of a strong shock interacting with a radiative 
cloud.  \citet{Mellema02} showed that the effects of a shock on 
such a cloud are very different from the non-radiative case.  
Instead of re-expanding and quickly diffusing into the background gas,
the compressed cloud instead breaks 
up into numerous dense, cold fragments.  These fragments survive for many 
dynamical timescales and are presumably the precursors 
to star formation.

In this work, we extend previous numerical studies by considering two
primary cooling models.  The first model uses an equilibrium cooling
curve with a tunable metallicity, set to solar in this work.  
As such, this model provides
a reasonable approximation for cooling in metal-enriched clouds 
around low-redshift (FRI) radio galaxies.
The second model solves the full non-equilibrium chemistry for a
primordial gas mixture.  This model is very well suited to address
jet-induced star formation in pristine clouds around high-redshift (FRII) 
galaxies.  With both models we explore a substantial subset of the
parameter space over which shock-induced 
star formation may apply.  We proceed in \S~\ref{sec:shockcloud} by discussing
the relevant physical processes, the timescales over which they act, and the
limitations of the present study.  The numerical methods and models
employed in the current
work are discussed in \S~\ref{sec:method}, and the results of the models are
presented in \S~\ref{sec:results}.  Further implications for star
formation are discussed in \S~\ref{sec:implications}.  Our conclusions are 
briefly recapped in \S~\ref{sec:conclusions}.

\section{Shock-Cloud Interactions}
\label{sec:shockcloud}

The basic idea of shock-induced star-formation is relatively 
simple.  A strong shock passes through a clumpy medium,
triggering many smaller-scale compressive shocks in overdense clumps.
These shocks increase the density inside the clumps and 
make it possible for them to radiate
more efficiently.  If the radiative efficiency of the gas has a sufficiently
shallow dependence upon the temperature, then radiative emissions 
are able to
cool the gas rapidly, in a runaway process, producing even higher 
densities as the
cooling gas attempts to re-attain pressure balance with the surrounding medium
\citep{Field65, MurrayLin89}.
Both the reduction in temperature and the increase in density act to reduce the
Jeans mass, above which gravitational forces become important. Any clump that
was initially close to this instability limit will be pushed over the 
edge and forced into gravitational collapse.  Thus, the passing of a 
shock through a clumpy medium may trigger a burst of star formation.
Several processes, acting on different timescales as discussed below, govern
whether or not cooling and star formation are able to proceed.

\subsection{Jet-Cloud Collision}
\label{sec:jetcloud}

In this work we are primarily interested in the interactions occurring 
over a relatively small region of a much larger intergalactic environment.  
However, since this work is being presented in the context of jet-induced 
star formation we wish to first motivate our treatment of the problem 
as either a single or a few overdense clumps being overrun by a planar
shock.  The presumption that the intergalactic medium is clumpy and 
multi-phase with a range of densities and temperatures 
is consistent with observational and theoretical 
studies of such objects as high-pressure ($10^5$ K cm$^{-3}$) 
cooling flow clusters \citep{Ferland02} 
and the giant emission-line halos found around distant radio galaxies 
\citep{Reuland03}.  The planar-shock assumption is also consistent, 
although the specific relation between the shock and the jet may 
vary somewhat from system to system.

For luminous FRII galaxies, with their powerful relativistic jets, 
it is unlikely that star formation will occur within the stream 
of the jet itself.  More likely, star formation will proceed 
within a cocoon of shock gas around the jet 
\citep{Begelman89,DeYoung89,Rees89}.  
Because of the very large sizes of these cocoons, the triggering 
shocks will appear to be nearly planar over the length scales 
considered in this study.

For low-luminosity FRI galaxies, the jets have a lower Mach number 
and appear turbulent; no hotspots or cocoons form \citep{bic84}.  
Nevertheless, such jets may still 
trigger potential star-forming shocks through collisions with 
nearby clouds, as illustrated in the following numerical 
simulation.  This simulation is performed using
the Cosmos code described in \S~\ref{sec:method} below.  Our parameters 
are chosen to closely resemble the conditions of Minkowski's Object, 
the small starburst system at the end of the radio jet emanating 
from the radio galaxy NGC~541.  Specifically, 
we set up an elliptical cloud with a major axis of 10 kpc, a minor 
axis of 5 kpc, a temperature $T=10^6$ K, and density 
$n_{cl} = 0.1$ cm$^{-3}$.  
This corresponds to an 
initial cloud mass of $\approx 10^9 M_\odot$.

The radio source associated with NGC~541 is of relatively low luminosity
and exhibits a FRI-type morphology \citep{Fanaroff74}. We, therefore, 
used the detailed radio and X-ray study of the proto-typical FRI-type
radio galaxy 3C31 by \citet{Laing02} to estimate plausible jet
parameters near Minkowski's Object, at $\sim 15$ kpc from the NGC~541 AGN.
For our numerical simulations, 
we assume that the long axis of the cloud is aligned with a jet of high
velocity ($9 \times 10^4$ km s$^{-1}$), low density ($10^{-4}$ cm$^{-3}$), 
hot ($10^9$ K) gas flowing onto 
our computational grid.  
The diameter of the jet nozzle is equal to half the diameter of the
cloud along the minor axis.  The assumption of an elliptical cloud 
is not essential to this discussion, although it can be easily 
motivated by tidal stretching from the nearby galaxies.  
It is also not essential that the jet be aligned with the long axis 
of the cloud, although this is convenient for numerical purposes.

As shown in Figure \ref{fig:jetcloud}
the jet-cloud collision triggers a nearly planar shock down the long axis
of the cloud.  As the bow shock from the jet wraps around the cloud,
it also triggers shocks along the sides of the cloud.  A similar shock
structure may explain the filamentary nature of the star-forming region in
Minkowski's Object.  Note the similarities between the next-to-last 
frame of the
numerical simulations and the observations as illustrated 
in Figure \ref{fig:MO}.

We stress that this simulation is only intended to motivate our 
subsequent treatment of the problem at much smaller physical scales.  
This simulation is unable to resolve the many moderately dense 
($\sim 10$ cm$^{-3}$), warm ($10^4$ K) sub-kiloparsec clouds presumed 
to be interspersed within the larger cloud simulated here.  
Further, the jet is not fully evolved when it hits the cloud, although 
this may not be critical since even a fully-evolved jet would trigger 
shocks in any clouds that crossed its path.

\subsection{Timescales}
\label{sec:timescales}

The initial conditions for all of our subsequent calculations consist of a
spherical cloud of radius $R_{cl}$ and density 
$\rho_{cl}=m_H n_{cl}$ in initial pressure equilibrium with a background
gas at temperature $T_{b,i}$ and density
$\rho_{b,i}=m_H n_{b,i} = \rho_{cl}/\chi < \rho_{cl}$, 
where $\chi>1$ is the density ratio between the cloud and background gases.
Given this simplified geometry and parameterization, a number
of useful dynamical timescales can be estimated analytically in order
to weigh the relative importance of various physical effects
including:  shock crossing, shock compression, instability
growth, cooling, freefall, magnetic acceleration,
and thermal conduction. 

The first timescale of interest is the shock-passage time.  
A planar shock of velocity 
$v_{sh,b}$ measured in the background gas, passes over the cloud in a time 
\begin{equation}
t_{sp} = \frac{2 R_{cl}}{v_{sh,b}} ~.
\end{equation}
The gas in the cloud will react to the shock on roughly a sound-crossing 
time 
\begin{equation}
t_{sc} = \frac{2 R_{cl}}{c_{s,cl}} ~,
\end{equation}
where $c_{s,cl}$ is the initial isothermal sound speed of the gas in the 
cloud.  Provided $c_{s,cl} \ll v_{sh,b}$, the cloud will rapidly
find itself 
in an overpressure region and a nearly spherical shock will be driven into
the cloud.  The velocity of this shock can be estimated by 
assuming pressure equilibrium between the post-shock cloud material 
and the shocked background region.  If the 
shock in the background is strong 
($\mathcal{M} = v_{sh,b}/c_{s,b} \gg 1$), then the 
post-shock pressure is approximately equal to $\rho_{b,i} v_{sh,b}^2$.  
Similarly, the 
post-shock pressure in the cloud is of order $\rho_{cl} v_{sh,cl}^2$.  
Equating these, we estimate the velocity of the shock in the cloud as 
\begin{equation}
v_{sh,cl} \simeq \left( \frac{\rho_{b,i}}{\rho_{cl}} \right)^{1/2} v_{sh,b} 
= \frac{v_{sh,b}}{\chi^{1/2}} ~.
\end{equation}
[More rigorous estimates are presented in \citet{Sgro75} and \citet{Klein94}.]  
This shock will compress the cloud on a timescale of approximately 
\begin{equation}
t_{cc} = \frac{R_{cl}}{v_{sh,cl}} \simeq \chi^{1/2} \frac{R_{cl}}{v_{sh,b}} ~.
\end{equation}
This is an important dynamical timescale for the cloud.

After the shock 
has passed over it, the cloud becomes susceptible to both 
Rayleigh-Taylor and Kelvin-Helmholtz instabilities.  The growth of 
the Rayleigh-Taylor instability is due to the acceleration of the 
cloud by the post-shock background gas.  The acceleration timescale 
is the amount of time it takes to accelerate the cloud gas to the velocity 
of the post-shock background.  If we consider the momentum transfer from 
a column of post-shock background gas with 
velocity $v_{b,ps}$ and cross-sectional area $\pi R_{cl}^2$ to a 
spherical cloud, then the acceleration timescale 
is given roughly by $t_{acc} \approx \chi R_{cl}/(3 v_{b,ps})$, 
where we have assumed a mean post-shock force acting
on the cloud of $\pi R_{cl}^2\rho_{b,ps} v_{b,ps}^2$ and used 
the shock-jump condition $\rho_{b,ps} = 4 \rho_{b,i}$ with 
an adiabatic index of $\Gamma=5/3$.
The shock-jump 
conditions also give us $v_{b,ps} = 3v_{sh,b}/4$.  
Thus, the acceleration timescale 
can be rewritten as 
\begin{equation}
t_{acc} \approx \chi \frac{R_{cl}}{v_{sh,b}} \approx  \chi^{1/2} t_{cc} ~,
\end{equation}
where we have dropped a numerical factor of order unity.  
This gives an acceleration of 
$g \sim v_{sh,b}/t_{acc} \sim R_{cl}/t_{cc}^2$, corresponding to a 
Rayleigh-Taylor growth time of $t_{RT}^{-1} \simeq (gk)^{1/2}$ 
for a perturbation of wavenumber $k$ \citep{Chandra61}.  
Written in terms of the 
cloud-compression timescale, $t_{cc}$, the Rayleigh-Taylor 
growth time is 
\begin{equation}
t_{RT} \sim \frac{t_{cc}}{(k R_{cl})^{1/2}} ~.
\end{equation}
Although the shortest wavelengths ($\lambda\ll R_{cl}$) have the fastest
growth rates, they also saturate rapidly when their amplitudes
$A\approx\lambda\ll R_{cl}$.  Wavelengths corresponding to
$k R_{cl} \sim 1$ are therefore the most disruptive.  Hence, 
Rayleigh-Taylor instabilities will break up the cloud over a timescale 
comparable to the cloud compression.

For $\chi \gg 1$ 
the timescale for growth of the Kelvin-Helmholtz instability is 
$t_{KH}^{-1} = k v_{rel}/\chi^{1/2}$ \citep{Chandra61}, 
where $v_{rel}$ is the relative 
velocity between the post-shock background and the cloud.  
Since the 
cloud accelerates rather slowly for $\chi \gg 1$, the 
relative velocity  
is approximately equal to the post-shock velocity of the background 
gas $v_{b,ps} = 3v_{sh,b}/4$.  
Again, in terms of the cloud-compression timescale $t_{cc}$ ,
we have
\begin{equation}
t_{KH} \sim \frac{t_{cc}}{(k R_{cl})} ~.
\end{equation}
The Kelvin-Helmholtz instability will thus also act on a 
cloud-compression timescale.

The timescales above have been derived assuming adiabatic evolution of
the gas and would be modified somewhat if radiative cooling were taken 
into account.  
However, our primary purpose is to roughly characterize 
some general timescales for these clouds, 
for which the values derived above are more
than adequate.

If the cloud is not able to cool radiatively, then the above results 
suggest that the cloud will be destroyed over a time comparable to the 
cloud-compression timescale.  This has been shown to indeed be the case 
in numerous 
hydrodynamic simulations \citep[e.g.][]{Klein94,Poludnenko02}.  
Destruction is enhanced by the fact that, 
after the compressive shock passes through itself and exits at the 
opposite surface of the cloud, it triggers a rarefaction 
wave traveling back into the cloud. This causes the cloud to re-expand
and increases the pressure contrast between the compressed cloud and
the post-shock external gas, effectively accelerating the growth of
destructive instabilities.

However, if the cooling timescale is short 
compared to the cloud-compression timescale ($t_{cool} \ll t_{cc}$), 
a very different outcome is 
reached, as was shown in \citet{Mellema02}.  
Cooling instabilities in the post-shock cloud material cause it 
to collapse into a very thin, cold dense shell behind the shock.  
This greatly enhances the density contrast between the post-shock 
cloud and background gases, which slows the growth of destructive 
instabilities.  
We can estimate the cooling 
timescale as 
\begin{equation}
t_{cool} = \frac{1.5 n_{cl,ps} k_B T_{cl,ps}}{(n_{cl,ps})^2 \Lambda(T)} ~,
\end{equation}
where the numerator is the energy density of the gas and 
the denominator is the volume emissivity.  
To get a general estimate of the cooling timescale, 
we use the approximate cooling rate, 
$\Lambda = 1.33 \times 10^{-19} T^{-1/2}$ erg cm$^3$ s$^{-1}$ 
from \citet{Kahn76}, appropriate for gas temperatures ranging from 
$5 \times 10^4 - 5 \times 10^7$ K.  Applying the shock-jump conditions,
the cooling timescale can be rewritten as
\begin{equation}
t_{cool} = C \frac{v_{sh,b}^3}{\chi^{3/2} \rho_{cl}} ~,
\end{equation}
where $C = 7.0 \times 10^{-35}$ g cm$^{-6}$ s$^4$.  In this study we 
will predominantly explore the cooling-dominated regime 
$t_{cool}<0.1t_{cc}$.  (A version of this condition is given in 
\citet{Mellema02}, Eq. 1.)  
For our typical data with cloud radius $R_{cl}=100$ pc and 
density ratio $\chi=10^3$, we find that cooling 
will generally govern evolution for moderate cloud densities and 
shock velocities. However, sufficiently high shock velocities
can suppress the effect of cooling over the cloud destruction time
(e.g., for shock velocities $v_{sh,b}>10^4$ km s$^{-1}$ 
with $n_{cl}\le1$ cm$^{-3}$). Cooling is also negligible at
sufficiently low densities 
(e.g., clouds with $n_{cl}<10^{-4}$ cm$^{-3}$ do not cool
for $v_{sh,b}\ge10^3$ km s$^{-1}$).

\subsection{Limitations}
\label{sec:limitations}

Our models are subject to some limitations. Due to resolution requirements
and computational limitations, the models have been studied in two-dimensional,
Cartesian geometry, and so the clouds represent slices through infinite
cylinders.  As can be seen in the models presented below, the initial
compressions are highly symmetric, and so the additional convergence expected
in three-dimensional models might well lead to stronger compressions and
enhanced cooling, especially for the primordial clouds which rely upon
the formation of H$_2$.  Three-dimensional simulations would also provide
an additional degree of freedom for fragmentation through dynamical
instabilities.  Such models might, therefore, be expected to lead to the
formation of greater numbers of fragments than seen below.

In all but one of our models, self-gravity is neglected, due to the geometry;
self-gravity in cylindrical symmetry is noticeably different than in 
spherical symmetry.  The relevant timescale for self-gravity is the free-fall
timescale of the gas, $t_{ff} = (3\pi/32G\rho)^{1/2}$. 
The initial overdense clumps considered here 
are specifically chosen not to be self-gravitating, and the corresponding 
freefall time is long compared to other relevant timescales.  However, as the
clouds compress, the local free-fall timescale becomes significantly
shorter.  For most of the runs, self-gravity is just becoming important
around the time we stop the simulations.  In order to explore the effects of
self-gravity, we include one run which solves for the self-gravity of
the cloud.  The results of that model indicate that the inclusion of
self-gravity would likely enhance compressions and cooling in all of the 
models.  

Examining three dimensional models with self-gravity, at
comparable resolution to that used in the models
presented here, will require the implementation of adaptive mesh
refinement to concentrate resolution only over those regions
which form dense fragments. That capability is currently
being added to the code described below, 
and results will be presented in future work.

If the background medium is magnetized, then magnetic field lines 
can become trapped in deformations on the surface of the cloud.  
As these field lines are stretched, the magnetic pressure along 
the leading edge of the cloud can increase enough to accelerate
the disruption of the cloud through the Rayleigh-Taylor 
instability \citep{Gregori99}.  We can estimate the importance of 
this effect by accounting for the increased acceleration of the 
cloud due to magnetic pressure \citep{Gregori99}
\begin{equation}
g = \frac{\rho_{b,i} v_{sh,b}^2 + P_B}{R_{cl} \rho_{cl}} ~,
\end{equation}
where $P_B = B^2/8\pi \equiv P_{b,ps}/\beta$.  The Rayleigh-Taylor 
growth time $t_{RT}^{-1}\simeq(gk)^{1/2}$ now becomes
\begin{equation}
t_{RT} \sim \frac{t_{cc}}{\{[1+3/(4\beta)]kR_{cl}\}^{1/2}} ~.
\end{equation}
From this estimate we see that, even for strong magnetic fields 
($\beta \lesssim 10$), the Rayleigh-Taylor growth time remains 
comparable to the cloud-compression timescale and hence longer 
than the cooling timescales for most of the runs considered.

Tangled magnetic fields within the clouds would also act to resist
compression, potentially reducing cooling, and enhancing cloud destruction
by shocks.  We are currently working to add MHD capabilities to our
numerical code, and shall examine the effects of realistic magnetic field
strengths in future work.

If thermal conduction is important then the cloud will evaporate 
into the hot background medium.  The rate at which the cloud evaporates 
can be written \citep{Klein94}
\begin{equation}
\dot{M}_{cl} = 4\pi R_{cl}^2 \rho c_s F(\sigma_0) ~,
\end{equation}
where $\sigma_0$ is the saturation parameter and $F(\sigma_0)$ is 
a dimensionless quantity of order unity.  If we define 
an ablation timescale $t_{ab} \equiv M_{cl}/\dot{M}_{cl}$, then in 
the post-shock environment 
\begin{equation}
t_{ab} \approx \frac{\chi^{1/2} t_{cc}}{7 F(\sigma_0)} ~.
\end{equation}
We see that for $\chi \gtrsim 10^2$, the ablation timescale will be 
comparable to or longer than the compression timescale.  Thus, for 
the parameters considered in this study, thermal conduction can safely 
be ignored.

\section{Numerical Methods}
\label{sec:method}

The numerical calculations discussed below have been computed using Cosmos, a
massively parallel,
multidimensional, radiation-chemo-hydrodynamics code designed 
for both Newtonian and relativistic flows developed at Lawrence
Livermore National Laboratory.  The relativistic capabilities and
tests of Cosmos are discussed in \citet{AF03}.  Tests of the Newtonian
hydrodynamics options and of the microphysics relevant to the
current work are presented in \citet{AFM03}, 
and so we shall not discuss those in detail here.
The calculations are carried out on a fixed, two-dimensional Cartesian
($x$,$y$) grid, implying that the simulated clouds are
cylindrical rather than spherical.  This limitation is 
currently necessary in order to maintain sufficient resolution 
to follow the fragmentation of the clouds.

\subsection{Cooling Models}

All of the remaining results presented here were performed at a 
fixed spatial resolution of 
$0.5 \times 0.5$ pc per zone.  There are 200 zones 
across the initial radius of the cloud, consistent with the resolution
requirements suggested by \citet{Klein94}.  We note, however, that
in the presence of cooling, which leads to even higher densities,
the resolution requirements become even more stringent.  We find that
we are only able to 
reliably follow the fragmentation of the cloud for slightly longer 
than the cloud-compression timescale, $t_{cc}$.  Beyond that point, 
further compression of the cloud is prevented by numerical 
resolution rather than any physical mechanism.  
We have allowed a few of our runs to evolve for longer times 
(on the order of a few $t_{cc}$) 
to confirm that the cold, dense cloud fragments which form 
are relatively long 
lived, consistent with the findings of \citet{Mellema02}.  
Further study of the evolution 
of these dense fragments will require finer resolution or an 
adaptive mesh.

\subsubsection{Equilibrium Cooling Curve - Low Redshift Systems}
\label{sec:equil}

Two radiative cooling and heating models are considered in this 
study.  In the first model, appropriate for enriched clouds around 
low-redshift galaxies, local cooling is given by the 
following cooling function:
\begin{equation}
\Lambda(T,n) = \left[\sum_i \dot{e}_i(T) (f_I n)^2 + \dot{e}_{M}(T)
f_M n^2 + J n \right]
    \times \left\{ \begin{array}{ll}
            \exp[(T-T_{min})/\delta T] & \mbox{if} ~ T \le T_{min} , \\
            1 & \mbox{otherwise} ,
            \end{array}
           \right.
\label{eqn:coolingnochem}
\end{equation}
based in part on an
equilibrium (hydrogen recombination and collisional excitation) cooling curve.
Here 
$\dot{e}_i$ is the cooling rate from hydrogen and helium lines,
$\dot{e}_M$ is the temperature-dependent cooling rate from metals
(including carbon, oxygen, neon, and iron lines, assuming solar metallicity),
$f_M$ is a weighting or tracer function for metal cooling taken to be unity,
$J$ is a generic background heating term, 
$f_I$ is an estimate of the ionization
fraction, defined as $\mbox{min}(1,~\mbox{max}(0,~(T_{eV} - T_c)/3) )$
with $T_c=1 eV$ to match roughly the expected upper and lower bounds in 
a mostly hydrogen gas equilibrium model, $T = (\Gamma-1)
e\mu /(k_B n (1+f_I))$ is the gas temperature in
Kelvin, $n$ is the number density of the gas, 
$e$ is the internal energy density
of the gas, $\mu$ is the mean molecular weight, assumed to be unity, 
and $k_B$ is Boltzmann's constant.  
The exponential, with width $\delta T = 1$ K, 
is introduced to suppress cooling below 
$T_{min}=10$ K.  
The cooling rate for metals, $\dot{e}_M$, is extended to
low temperatures ($\sim 10$ K) using the 
curves of \citet{Dalgarno72} in the low ionization limit.
The background heating ($J$) is set to 
initially balance the cooling inside the cloud and remains fixed 
throughout the evolution, consistent with the treatment of 
\citet{Mellema02}.

\subsubsection{Non-equilibrium Primordial Chemistry - High Redshift Systems}
\label{sec:chem}

The second cooling model applies when the chemistry is solved 
dynamically with the full non-equilibrium equations as described in 
\citet{AFM03}.  
Currently, nine atomic and molecular species are 
included in the chemistry model: 
\ion{H}{1}, \ion{H}{2}, \ion{He}{1}, \ion{He}{2}, \ion{He}{3}, 
$e^-$, H$^-$, H$_2$, H$_2^+$.
A total of 27 gas-phase chemical reactions are included in the full
network.  As such, this model is useful for considering pristine 
clouds around high redshift galaxies.  
The various ionization states 
and concentration densities $n_{i}$ of each species
are calculated from the time-dependent chemistry 
equations using a sequential backwards
differencing scheme \citep{Anninos97} and used explicitly in the
cooling function as
\begin{equation}
\Lambda(T,n^{[m]}) =
    \sum_{i=1}^{N_{s}} \sum_{j=1}^{N_{s}}  \dot{e}_{ij}(T) n^{[i]} n^{[j]}
    + \sum_{i=1}^{N_{s}} J_i n^{[i]}
    + \dot{e}_M(T) f_M n^2
    ,  \label{eqn:coolingchem}
\end{equation}
where $\dot{e}_{ij}(T)$ are
the cooling rates from 2-body interactions between species
$i$ and $j$, and $J_i$ represents frequency-integrated photoionization
and dissociation heating.
The equation of state for temperature in this case is given by
$T = (\Gamma-1) e/(k_B {\sum_i n_i})$.
We account for a total of seven different cooling and heating mechanisms:
collisional-excitation,
collisional-ionization,
recombination,
bremsstrahlung,
metal-line cooling (dominantly carbon, oxygen, neon, and iron),
molecular-hydrogen cooling, and photoionization heating.
The models with non-equilibrium chemistry
also include photoionization of H and photodissociation of H$_2$
by free-streaming radiation fields.  For most models, the
photoionization rate is taken to be the appropriate rate expected from
cosmic UV background radiation at low redshift \citep{Bechtold87}, while
the photodissociation rate is taken to be the value applicable to the
local interstellar medium \citep{Spaans97}.  We also examine excursions
by an order of magnitude from those values.

The essential difference between this cooling model from the
equilibrium model of section \ref{sec:equil} is that the atomic
and molecular reactions are properly taken into account in order
to resolve temporal phase differences between the cooling and recombination
times. This allows us to more accurately predict the concentration of
residual free electrons and ions from the slower recombination
processes, particularly at and below the hydrogen-line
cooling curves which dominate cooling down to about $10^4$K. 
These residual electrons are captured
predominately by neutral hydrogen to form H$^-$, which
subsequently produces hydrogen molecules through collisional
interactions with other hydrogen atoms.
Cooling below the hydrogen Lyman-$\alpha$ line edge is achieved
through the excitation of the vibrational/rotational modes of
hydrogen molecules, provided H$_2$ forms in sufficient abundance.
For molecular line excitations, we use the cooling
function of \citet{LS83}, from which we can expect
additional cooling down to about a couple hundred degrees Kelvin
for our typical cloud parameters and limited grid
resolution. Presumably the inclusion
of dust grain physics or deuterium and higher order chemistry
would contribute to further cooling. We will investigate these
effects in future papers, but in this present work, we
do not expect to achieve the same level of cooling in the chemistry model as
in the equilibrium model, since the latter 
includes low temperature ($<100$K) cooling from metals.

In the non-equilibrium chemistry model,
we also account for the self-shielding of the cloud against
photoionizing and photodissociating background fields.  We do this by 
calculating an optical depth
\begin{equation}
\tau^{j}_i(\ell) = \int_{\ell_0}^{\ell} \sigma_i n_i d\ell^{\prime}
\end{equation}
for the two species of interest: \ion{H}{1} and H$_2$.  The limits of
integration run from the edge of the grid to points within the cloud,
along the four cardinal directions (see below).  The cross sections
used are
$\sigma_{\mathrm{HI}} = 6.3 \times 10^{-18}$ cm$^2$ \citep{Osterbrock89} 
and 
$\sigma_{\mathrm{H}_2} = 5.2 \times 10^{-18}$ cm$^2$ \citep{Hollenbach71}.  
These are the peak values of the respective cross sections.  
The photoionization 
cross section of H drops off rapidly above the ionization threshold, and
so photoionization is dominated by photons near 13.6~eV, making that the
most appropriate value for our approximate treatment.  Photodissociation
of H$_2$ occurs for a narrow range of energies below 13.6~eV, and so the
cross section can be considered to be constant.  For high column
densities, the wings of the absorption become important, for which the
cross section would be less than used above.  Our
models do not achieve extremely high cross sections, and so we use the
larger value of $\sigma_{\mathrm{H}_2}$.
The optical depth calculation is initiated from each of the outer
boundary cells of the computational grid and integrated along
the Cartesian axes perpendicular to the boundary cell faces.
For any interior zone located at position $x$, the optical depth is taken
to be the minimum of all optical depth integrations which
intersect that cell position from every exterior boundary. Hence we take
\begin{equation}
\tau_i(x) = \mbox{min}\left[  \tau^{-x}_i(\ell), ~\tau^{x}_i(\ell),
                             ~\tau^{-y}_i(\ell), ~\tau^{y}_i(\ell) \right] ,
\end{equation}
where the superscript indicates the line-of-sight direction 
of integration intersecting at a cell center $x$.  Any symmetry 
boundaries, such as the lower $y$-boundary in this work, are not 
included in this calculation, since the optical depth in this 
direction would necessarily be higher.  
The appropriate reaction rates and heating coefficient are modified
to account for the integrated absorption as 
\begin{eqnarray}
k20 & \rightarrow & k20 \exp(-\tau_{\mathrm{HI}}) ~, \nonumber \\
k27 & \rightarrow & k27 \exp(-\tau_{\mathrm{H}_2}) ~, \nonumber \\
J_{\mathrm{HI}} & \rightarrow & J_{\mathrm{HI}} \exp(-\tau_{\mathrm{HI}}) ~,
\end{eqnarray}
following the notation of \citet{AFM03}.

\subsection{Parameter Space}

For all of our calculations we have considered $R_{cl} = 100$ pc, 
$T_{cl} = 10^4$ K, and $\chi = 10^3$ to 
be fixed.  Assuming that the background is initially in pressure 
equilibrium with the cloud, we are left with only a two-dimensional 
parameter space to explore, $n_{cl}$ vs. $v_{sh,b}$, for the 
equilibrium cooling model.  For the non-equilibrium chemistry 
model we will also consider additional parameters related 
specifically to that cooling model.  For this combination of fixed
parameters, the sound-crossing time for the cloud in every case 
is $t_{sc}=17$ Myr.

Before conducting any simulations, 
we first wish to narrow the parameter space to be studied.  Since we are 
primarily interested in shock-induced star formation, we want 
to concentrate on clouds that are able to cool efficiently during 
their compression.  We can define an approximate cooling-dominated 
regime from the timescales for cooling and cloud-compression: 
$t_{cool} < 0.1 t_{cc}$.  Rewriting this in terms of the 
variables used in our numerical study, we find that cooling dominates
provided that
\begin{equation}
\rcl > 1.2 \times 10^{-4} \denratio^{-2} \vshb^4 \ncl^{-1} ~.
\end{equation}
(A similar condition is given in \citet{Mellema02}, Eq. 1.)  
In the context of shocks driven by jets from AGN, 
this illustrates that cooling can dominate over much of the parameter 
space of interest.  The dashed line in Figure \ref{fig:equil} 
separates the cooling-dominated regions from the non-cooling 
for a fixed cloud radius of $R_{cl}=100$ pc and density ratio $\chi=10^3$.

We arrive at another constraint by considering the free-fall timescale 
of the radiatively cooled gas, $\tau_{ff}$, which represents the shortest
possible timescale on which stars can form within the cold gas.  It 
also represents a typical spread over which star formation shall occur.
As the first massive stars form, they provide an extra heating source in
the cloud that will likely shut off any further star formation
\citep{LinMurray00,DLM03}.  If we take 1~Myr as a typical star-formation
timescale, then any gas with a longer free-fall timescale is unlikely to
form stars efficiently.  Thus we define $t_{ff} \le 1$ Myr as our criterion
for efficient star-formation.  We can estimate 
the free-fall timescale $t_{ff} = (3\pi/32G\rho)^{1/2}$ by 
assuming that the cold cloud gas will eventually reach pressure 
equilibrium with the post-shock background.  With this assumption, the final 
density of the cloud material is 
\begin{equation}
\rho_{cl,f} \approx \frac{\mu m_H P_{b,ps}}{k_B T_f} ~.
\end{equation}
Using the shock-jump condition 
$P_{b,ps}\approx \rho_{b,i} v_{sh,b}^2 = (\rho_{cl} v_{sh,b}^2)/\chi$, 
we can write our star-formation criterion as  
\begin{equation}
\left( \frac{t_{ff}}{\mathrm{Myr}} \right) \le 
1.5 \denratio^{1/2} \ncl^{-1/2} \vshb^{-1} 
\left( \frac{T_f}{100 \mathrm{ K}} \right)^{1/2} ~.
\end{equation}
The dot-dashed line in Figure \ref{fig:equil} shows this
star-formation cut-off for specific values of 
$R_{cl}=100$ pc, $\chi=10^3$, $T_f = 100$ K, and $t_{ff}=1$ Myr.

Two additional constraints come from the requirements 
that the cloud not be Jean's unstable 
initially and that the shock velocity necessarily exceed the sound speed 
in the background gas.  The 
first requirement gives us a constraint relation between the 
radius, temperature, and density of the cloud
\begin{equation}
\ncl < 91 \rcl^{-2} \left( \frac{T_{cl}}{10^4 \mathrm{ K}} \right) ~.
\end{equation}
The horizontal solid line in Figure \ref{fig:equil} shows this 
stability limit for $R_{cl}=100$ pc and $T_{cl}=10^4$ K.
The second requirement can be written as a relation between the 
velocity of the shock and the temperature of the cloud
\begin{equation}
\vshb > 3.7 \times 10^2 \denratio^{1/2} 
\left( \frac{T_{cl}}{10^4 \mathrm{ K}} \right)^{1/2} ~.
\end{equation}
This constraint is easily met by all of the parameters considered in 
this study and is not shown in Figure \ref{fig:equil} (it 
would be a vertical line just off the left edge of the figure).

Thus we can divide our parameter space into five regions as shown 
in Figure \ref{fig:equil}.  Region I 
(gravitationally unstable) is not of interest in this study.  
In regions IV and V cooling is expected to be negligible, so these are 
not of much interest either.  Although cooling will be important in 
region III, the star-formation timescale is too long for the 
process to be very efficient.  This leaves us with region II as the 
sector of primary interest in this study of shock-induced star formation.
However, we also examine models in region~V in order to confirm our
above estimates of the location of the boundary between regions.  
The particular parameter pairs chosen for numerical study (marked with 
crosses in Figure \ref{fig:equil}) are discussed below.

\section{Results}
\label{sec:results}

Table \ref{tab:equil} summarizes the runs using the equilibrium 
cooling-curve model with solar metallicities.   We choose ten
combinations of $n_{cl}$ and $v_{sh,b}$ in order to explore the
parameter space of interest indicated in Figure~\ref{fig:equil}.
The Mach number, $\mathcal{M}$, corresponding to each value of 
$v_{sh,b}$ is also listed.  Finally, Table \ref{tab:equil} also 
includes the shock-passage ($t_{sp}$), cloud-compression ($t_{cc}$), 
and cooling 
($t_{cool}$) timescales for each run.  Notice that the shock-passage 
and cloud-compression timescales vary by less than an order 
of magnitude, whereas the cooling timescale 
varies by almost 5 orders of magnitude 
among the models considered.  This again arises from the very strong 
dependence of $t_{cool}$ on $v_{sh,b}$ and $n_{cl}$.

Table \ref{tab:chem} summarizes the runs using the non-equilibrium 
primordial chemistry model.  We choose a subset of six  parameter 
combinations from those considered with the equilibrium cooling-curve model.  
Table~\ref{tab:chem} also 
includes the shock-passage ($t_{sp}$) and cloud-compression ($t_{cc}$) 
timescales for each run.

\begin{figure}
%Fig 4 {Jet-2D-23_den.eps}
\epsscale{0.5}
\plotone{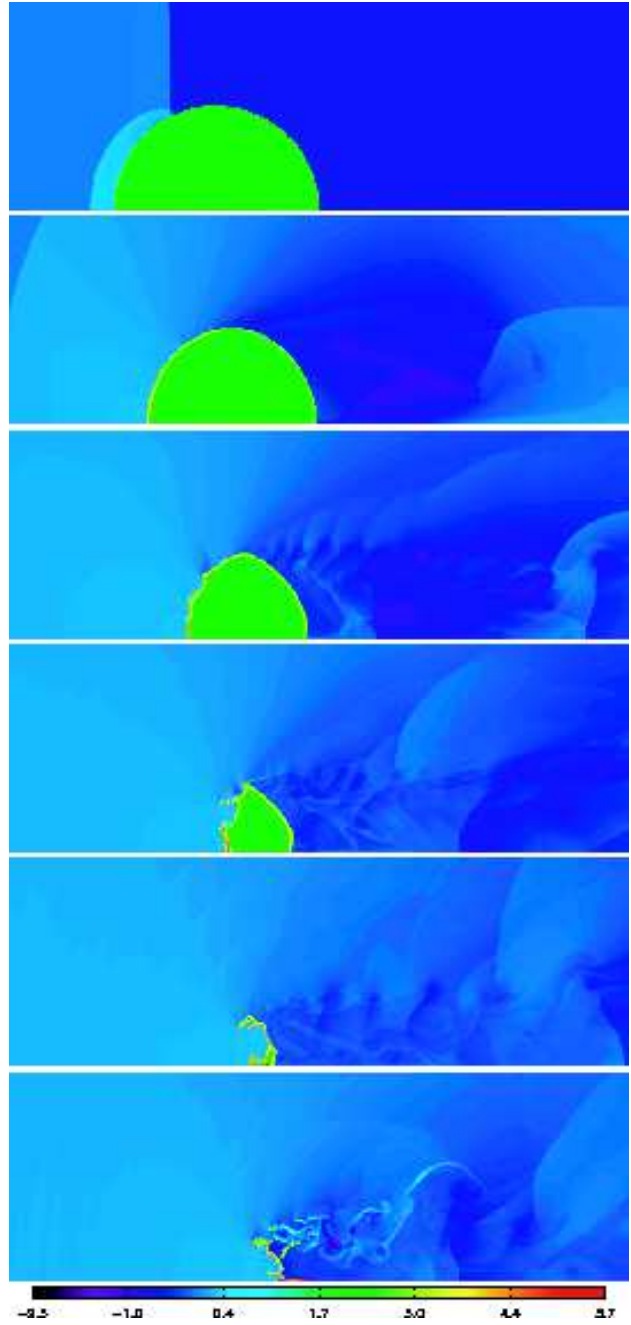}
\caption{Contour plots of the logarithm of gas density for model 
E3 at times $t=0.05$, 0.25, 0.5, 0.75, 1.0, and 1.25 $t_{cc}$.  The 
units for the density scale in this plot are $6.7 \times 10^{-26}$ 
g cm$^{-3}$, corresponding to an initial cloud density of 
$1.7 \times 10^{-23}$ g cm$^{-3}$.  Each frame shows the full 
computational grid, with physical dimensions of $600 \times 200$ pc.
}
\label{fig:modelE3}
\end{figure}

Figure \ref{fig:modelE3} illustrates a sequence of density contour plots 
taken from the simulation of model E3.   The parameters for this 
model were intentionally chosen to match those of run A in 
\citet{Mellema02}.  As expected, our results match theirs quite well.  
The strong cooling behind the compression shock in the cloud 
causes a very dense shell to form.  The density of this shell 
continues to increase as it follows the compression shock into 
the center of the cloud.  
Hydrodynamic and radiative instabilities cause the cold shell 
to fragment as it 
progresses toward the center of the cloud.  
Similar results are seen in all of our 
cooling-dominated runs.  As highlighted in 
\citet{Mellema02}, the fragmentation is a unique result of 
hydrodynamic simulations of shock-cloud interactions in the 
cooling-dominated regime  and is not observed in 
simulations that ignore radiative cooling \citep{Klein94}.  
In our simulations, the final densities of the fragments are 
often 3-4 orders of 
magnitude higher than the initial cloud density.  However, 
as mentioned above, these final densities are generally limited by numerical 
resolution and are likely to underestimate the true 
final densities in these fragments.  We are also unable to clearly 
resolve any interactions among either the main colliding shocks or
the cold dense filaments of gas at late times.

\subsection{Cooling Efficiency}
We attempt to quantify the efficiency
of the cooling processes in each of the runs presented.
In order to accomplish this, we track the gas inside the cloud 
using a tracer fluid which is passively advected
in the same manner as the density.  Throughout each run the 
tracer distribution reflects the distribution of original cloud 
material.  More importantly, it allows us to quantify how much of the 
initial cloud material cools below certain cutoff temperatures.  
Table~\ref{tab:cooleq} gives the percentages of cloud material that cool below 
$T = 1000$ K and $T=100$ K for the equilibrium cooling curve 
models.  The thermally
unstable regime of cooling ends between $100$ and $1000$~K, so the 
fraction of gas with $T<1000$~K indicates the relative 
importance of cooling to the
cloud evolution.  Extensive star formation is unlikely, however, unless the
gas is able to cool to $T<100$~K.  The ability of the gas to reach such
low temperatures in our equilibrium cooling curve models is, however, 
affected by numerical 
resolution and the neglection of self-gravity, 
both of which affect the peak density, and
hence the cooling efficiency of the gas.  
Therefore, the percentage of gas that cools to below 1000~K gives a
strong upper limit to the percentage that might form stars, while,
when well resolved, the amount that cools to below 100~K gives a more accurate 
measure.
These results are also represented 
graphically in Figure~\ref{fig:equil}.  
We see that the equilibrium cooling curve results 
agree quite well with the expected transition from the non-cooling 
to the cooling-dominated regime, shown by the dashed line.  

We also note from the results that the cooling process is generally
extremely efficient throughout the cloud.  In the most extreme case 
considered using the equilibrium cooling-curve (model E2), 77\% of the gas
in the cloud or $7.7 \times 10^5 M_\odot$ cools below 100 K.  Large
fractions of cool gas were seen in all of the other models within
region II.  The smallest value (21\%) occurs near the boundary between regions
II and V, while a more typical average seen for the models is approximately
50\%.  The primary exception is
in model E6b, in which three clouds were modeled (see below).

Table \ref{tab:coolch} reports these same percentages for the 
non-equilibrium chemistry models.  As expected, none of these 
models cool below 100 K (see \S \ref{sec:chem}), although many of 
them show substantial cooling below 1000 K.  
Cooling is also less efficient in these models than with the 
equilibrium cooling-curve.  
In the non-equilibrium chemistry models, 
cooling below $10^{4}$~K is due solely to H$_2$ emission.
While H$_2$ is an efficient coolant, the small fractions formed via gas-phase
reactions (Table~\ref{tab:coolch}) are insufficient to lead to cooling as
extensive as that due to metals, which dominate the equilibrium 
cooling curve models.  Based upon our numerical results, it 
appears the cooling timescales for the non-equilibrium chemistry 
models are roughly a factor of 10 longer than for the equilibrium 
cooling curve models using the same parameters.  
The decreased cooling efficiency leads
to a shift in the boundary between regions II and V towards higher densities
and lower shock speeds, as can be seen in Figure~\ref{fig:chem}.
Nevertheless, large fractions of
the gas contained within the clouds still cool to below 1000~K.  Given 
higher grid resolution, self-gravity, and an extended chemical
network including metals, each of which would be expected to
enhance the cloud densities and therefore the H$_2$ formation and cooling,
a substantial fraction of gas might also be expected to cool below 100~K.
The results indicate, therefore, that star formation may be extremely 
efficient in shocked clouds for which the cooling timescales are
sufficiently short.

\subsubsection{Self-Gravity}

The initial free-fall timescales for our models vary 
from $7.3 \times 10^6$ Myr for $n_{cl}=50$ cm$^{-3}$ (run E1) to 
$1.6 \times 10^8$ Myr for $n_{cl}=0.1$ cm$^{-3}$ (runs E8-E10).  
As explained above, our parameters were deliberately chosen such 
that the clouds would not be gravitationally unstable initially and 
that the free-fall timescale would be longer than the dynamical 
timescale (taken as the cloud compression time $t_{cc}$).  
Nevertheless, as we have seen, 
the density of the gas increases substantially in the radiative 
shell behind the compression shock.  Therefore the local free-fall time 
of this compressed gas will be much shorter than its initial 
value.  This, of course, is the basis of the 
shock-induced star-formation model.

For a fairly typical density increase of $10^4$, the local free-fall 
timescale will decrease by $10^2$.  In the coldest gas, the temperature
decreases by more than two orders of magnitude from its original value,
and so the cold, dense regions have Jeans masses reduced by five orders
of magnitude relative to the original cloud.  Thus, for the higher density 
runs (E1-E4), the self-gravity of the gas becomes dynamically important 
in the dense fragments behind the compression shock.  Although 
the Cosmos code used in this study includes an option for solving 
self-gravity, most of our runs did not utilize it.  In the
two-dimensional limit considered in the current study, we would be
solving the gravity of an infinite cylindrical cloud, which is notably
different than the self-gravity of a spheroidal cloud. Nevertheless, we
did conduct one run (E1b) with the self-gravity option turned on.  The
enhanced compression due to self-gravity did allow the gas to cool
more efficiently than the corresponding run without self-gravity (E1a).  As
a result, 69\% of the gas in model E1b cooled to below 100~K, as compared
to 57\% in model E1a.

\subsubsection{Multiple Clouds}

\begin{figure}
%Fig 6 {Jet-2D-47_den.eps}
\epsscale{0.75}
\plotone{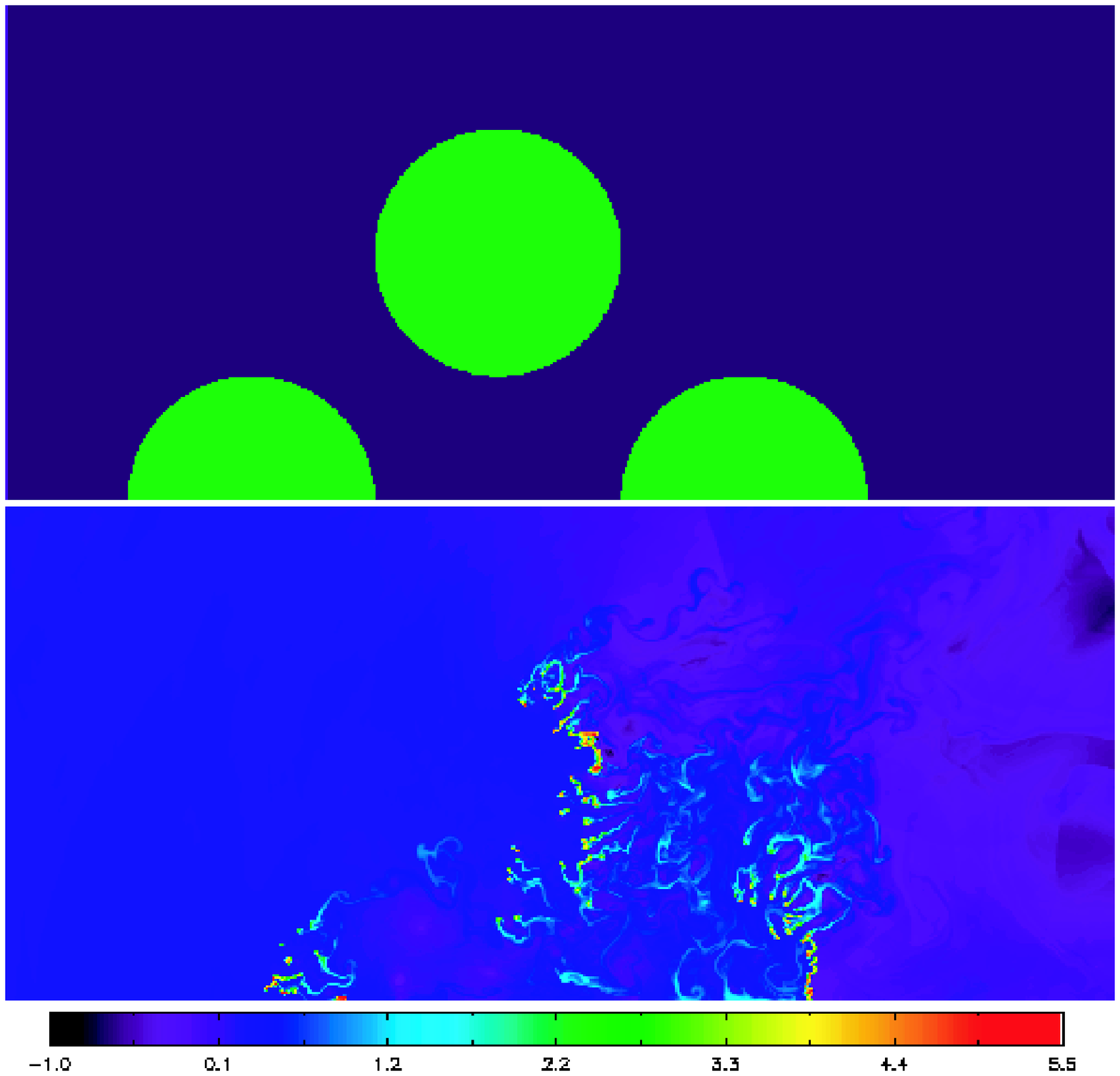}
\caption{Contour plot of the logarithm of gas density for model E6b 
at times $t=0$ and $1.25 t_{cc}$.  The units for the density scale 
in this plot are $6.7 \times 10^{-27}$ g cm$^{-3}$, 
corresponding to an initial cloud density of 
$1.7 \times 10^{-24}$ g cm$^{-3}$.  Each frame shows the full 
computational grid, with physical dimensions of $900 \times 400$ pc.
}
\label{fig:multi}
\end{figure}

For most of this study we consider the interaction of a planar 
shock with a single cloud.  However, our general notion of 
jet-induced star formation envisions the interaction of such a 
shock with a system of clouds within an inhomogeneous background.  
To highlight the possible effects of the interaction of multiple clouds, 
we have considered one run with 
a system of four clouds: two aligned perpendicular to the shock front,
and two aligned parallel to it. This set up 
allows us to achieve higher spatial resolution by imposing
symmetry boundary conditions and computing only half of the problem.
Each individual cloud has the same parameters as 
the single cloud in model E6a ($R_{cl} = 100$ pc, $T_{cl}=10^4$ K, 
and $n_{cl} = 1$ cm$^{-3}$).  The initial cloud configuration 
is shown in the top of Figure \ref{fig:multi}.
The size of the computational grid and the number of zones are 
scaled equally to maintain the same 
$0.5 \times 0.5$ pc per zone resolution as our other runs.

The interaction of these clouds results in less efficient cooling 
throughout all of the clouds, but particularly for those in the 
downwind direction.  The leading cloud, in fact, cools almost identically 
to the isolated cloud case.  The discrepancy in the amount of gas 
found below our 100 K threshold (9\% for the front cloud in run E6b 
versus 21\% for the isolated cloud in run E6a) 
disappears if we change the lower threshold to 200 K.  The amount of 
gas below that threshold is 28\% for both 
models.  This indicates 
that there is a significant amount of gas near 100 K in both models, 
but in the multi-cloud model most of it is just above this 
threshold.

The reduced amount of cold gas in the downwind clouds is likely due
to a combination of effects.  First, geometric screening of the primary shock 
results in less symmetric, and generally less efficient, 
compression for the downwind clouds.  The close proximity 
of the clouds in this simulation also subjects them to multiple reflected 
shocks from their neighbors which can reheat previously cooled gas.  
This may be 
enhanced in the present work by the symmetry imposed by the 
two-dimensional geometry.  In three dimensions, these reflected 
shocks would presumably be less focused.  Finally, consistent with 
\citet{Poludnenko02}, we find that the channel between the clouds acts 
as a de Lavalle nozel and accelerates the background gas.  Since both the 
Rayleigh-Taylor and Kelvin-Helmholtz instabilities scale as the 
inverse of the background velocity, this acceleration reduces the 
destruction time for the downwind clouds.  This effect may 
be reduced in three dimensions.

The results of this multi-cloud run suggest
that the efficiencies we quote for our single cloud runs should
generally be viewed as applicable to relatively isolated clouds.  We note,
however, that although reduced by a factor of a few relative to the 
single-cloud models, the fraction of cold gas is still significant in
the model with multiple clouds.  In future work we plan to present 
a more detailed study of the interactions of shocks with systems 
of radiating clouds.

\subsection{H$_2$ Formation}

\begin{figure}
%Fig 7 {Jet-2D-29_all.eps}
\epsscale{0.5}
\plotone{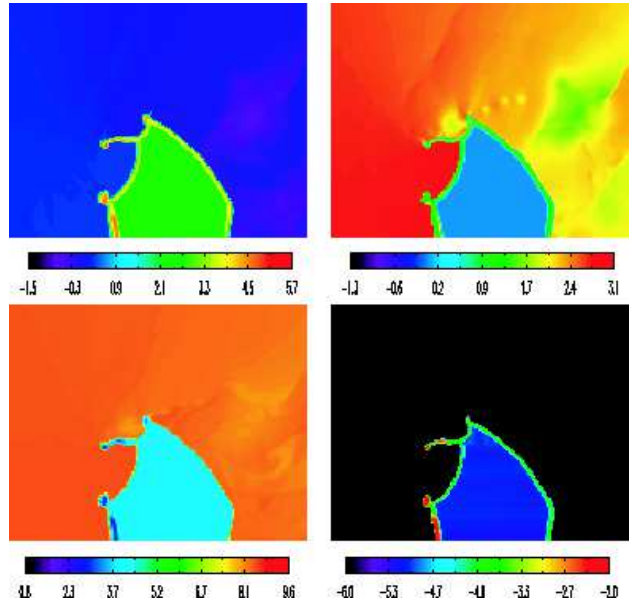}
\caption{Contour plot of the logarithm of gas density ({\it upper left}), 
gas pressure ({\it upper right}), temperature ({\it lower left}), and 
H$_2$ mass fraction ({\it lower right}) for 
model C1a at time $t=0.75 t_{cc}$.  The corresponding units are: 
$1.7 \times 10^{-23}$ g cm$^{-3}$ for density and 
$8.9 \times 10^{-12}$ g cm$^{-1}$ s$^{-2}$ for pressure.  
The temperature unit is Kelvin.  The H$_2$ mass fraction 
measures $\rho_{H_2}/\rho$.  The plots only show a small portion of 
the overall computational domain.
}
\label{fig:H2}
\end{figure}

Table \ref{tab:coolch} also lists the local peak mass fraction 
of H$_2$ [max($\rho_{H_2}/\rho$)] and the total H$_2$ mass fraction 
($M_{H_2}/M_{cl}$) for the non-equilibrium chemistry models, 
where $M_{H_2}$ is the total mass of H$_2$ at the end of each 
run ($t=1.25 t_{cc}$) and $M_{cl}$ is the initial cloud mass.  
The observed H$_2$ mass fractions are consistent with the results of 
\citet{Anninos97}.
Figure \ref{fig:H2} illustrates how the H$_2$ distribution, 
temperature, density, and pressure trace each other for model C1a.

\subsubsection{Chemistry Parameters}
Along with the six combinations of $n_{cl}$ and $v_{sh,b}$ studied 
with the non-equilibrium chemistry model, we also explore the 
dependence of our results on our treatments of photoionization and 
photodissociation.  Model C1a is our base model, comparable to the 
equilibrium-cooling model E3.  Models C1a and C2-C6 include photoionization, 
photodissociation, and integrated optical depth calculations.  
Model C1b ignores photoionization and 
photodissociation.  In the absence of these processes, there is 
also no need to calculate any optical depths.  
This run results in a slightly higher percentage 
of gas cooling below 1000 K, relative to model C1a, 
due to a higher mass fraction of H$_2$.  
Model C1c includes photoionization and photodissociation but does not 
include optical depth calculations.  The omission of self-shielding 
results in a 
dramatic reduction in the amount of gas which is able to cool 
below 1000 K and in the amount of H$_2$ that forms, 
illustrating the importance of self-shielding.  
Model C1d increases the strength of the external photoionization field 
by a factor of 10.  This produces a slight increase in the percentage of 
gas able to cool below 1000 K, due to the increased ionization fraction,
which, in turn, enhances the formation of H$_2$ via H$^-$.
Model C1e increases the H$_2$ photoreactive destruction rate 
coefficient ($k27$) by 
an order of magnitude.  This results in a reduction of about 14\% 
in the amount of gas cooling below 1000 K and a dramatic reduction 
in the final quantity of H$_2$.

\section{Implications for Cloud and Jet Evolution}
\label{sec:implications}

Our numerical studies have concentrated primarily on 
following the evolution of 
a single dense cloud being overrun by a planar shock traveling 
through a low density background.  This picture is 
intended to represent what is happening on a relatively small 
scale ($\sim 100$ pc) within a much larger region of interaction 
($\gtrsim 1$ kpc) between a jet-induced shock and an inhomogeneous 
medium.  We now 
generalize our results to predict some properties of jet-induced 
star-forming regions.

\subsection{Star Formation Rates}

We first wish to consider the star-formation rate for this process.  
To arrive at this, we need to estimate the rate at which the 
mass of gas in dense clumps is swept over by the shock.  This is given by 
\begin{equation}
\dot{M}_{cl} = f_{cl} \rho_{cl} v_{sh} A ~,
\end{equation}
where $f_{cl}$ is the volume filling factor of the clumps and 
$A$ is the surface area of the shock front.  The star-formation rate 
is the mass rate multiplied by the star-formation efficiency.  
Here it will help to consider a specific object, so we take 
Minkowski's Object.  From our simulations in \S \ref{sec:jetcloud}, 
we estimate the shock speed inside the cloud 
to be $\approx 3 \times 10^8$ cm s$^{-1}$.  
From \citet{vB85} we take 
$A \approx 20$ kpc$^2$, where we assume, for simplicity, that 
the cross-sectional area is constant.  
Then, for a single dense phase of 
$\rho_{cl} \approx 10^{-23}$ g cm$^{-3}$, we can match the 
observed star-formation rate of $\sim 0.3 M_\odot$ yr$^{-1}$ 
by assuming a volume 
filling factor of $f_{cl} \approx 0.03$ and 
a star-formation efficiency of 0.1\%.

\subsection{Acceleration of Clouds and Mass Loading}

We have already estimated the acceleration timescale of these 
clouds in \S \ref{sec:timescales}.  From this we can estimate 
their velocities as a function of time.  In 
terms of the typical parameters considered in this study, we get 
\begin{equation}
\left( \frac{v_{cl}}{\mathrm{km/s}} \right) 
\lesssim 20 \denratio^{-1} \rcl^{-1} \vshb^2
\left( \frac{t}{\mathrm{Myr}} \right) ~.
\end{equation}
This value seems to overestimate the cloud velocity by a factor of 
a few based upon our numerical results.  However, this is reasonable 
since the estimate assumes 
a constant cross-sectional area for the cloud.  
As we have seen, the cross-sectional area 
can be reduced significantly 
during compression.  The final conclusion is that, for the 
parameter space studied, these clouds accelerate 
very slowly.  We note that, for Minkowski's Object, this is in good 
agreement with observations, which show very little evidence for 
velocity gradients in the ionized gas downstream from the jet-cloud 
collision region. 
%CAN WE COMPARE THIS WITH THE VELOCITY OF THE FRAGMENTS FROM
%THE CODE OUTPUT? 
%HOW DOES THIS COMPARE WITH THE MEAN FLOW WITHOUT COOLING?
%DOES COOLING RADIATE AWAY ALOT OF KINETIC ENERGY?  ANSWERS:  AS I 
%NOTE ABOVE, THE ANALYTIC ESTIMATE AGREES FAIRLY WELL WITH OUR 
%NUMERICAL RESULTS.  THE VELOCITIES APPEAR TO BE SIMILAR FOR BOTH 
%RADIATIVE AND NON-RADIATIVE CLOUDS (WITHIN A FACTOR OF A COUPLE). 
%HOWEVER, THAT CONCLUSION IS BASED UPON 
%AN EYEBALL ANALYSIS OF THE X VELOCITY COMPONENT IN IDL AND LINE PLOTS.  
%BASED UPON THIS I WOULD GUESS THAT NOT A LOT OF KINETIC ENERGY IS 
%RADIATED AWAY, BUT I DON'T HAVE A DETAILED ANSWER.  IF YOU THINK IT'S 
%WORTH THE TIME I COULD TRY TO WORK OUT THE KINETIC ENERGY CONTENT OF THE 
%CLOUD GAS FOR A RADIATIVE AND NON-RADIATIVE CASE TO COMPARE.

Along with the potential of becoming starburst regions, the nearly 
stationary cloud fragments may also become significant sources of 
mass loading for the post-shock flow \citep{Hartquist88}.  
Mass loading is the feeding of material into the flow and 
could have a significant impact on its properties, 
including possibly causing it to transition to the 
transonic regime.  However, it is difficult to estimate the 
significance of mass loading from our simulations.  Over 
the evolution time of our models ($1.25 t_{cc}$), very little 
mass is lost from the clouds (typically $\lesssim$ 1\%).  This 
amount would certainly increase if longer evolution times were 
considered, but as we pointed out in \S \ref{sec:method} we can 
not reliably follow the evolution of these clouds beyond the current limits 
since our numerical resolution prevents further compression of the clouds.  
The question of mass loading is one we will return to in future 
studies.

\section{Conclusions}
\label{sec:conclusions}

\begin{figure}
%Fig 3 {paramequil_bw.eps}
\plotone{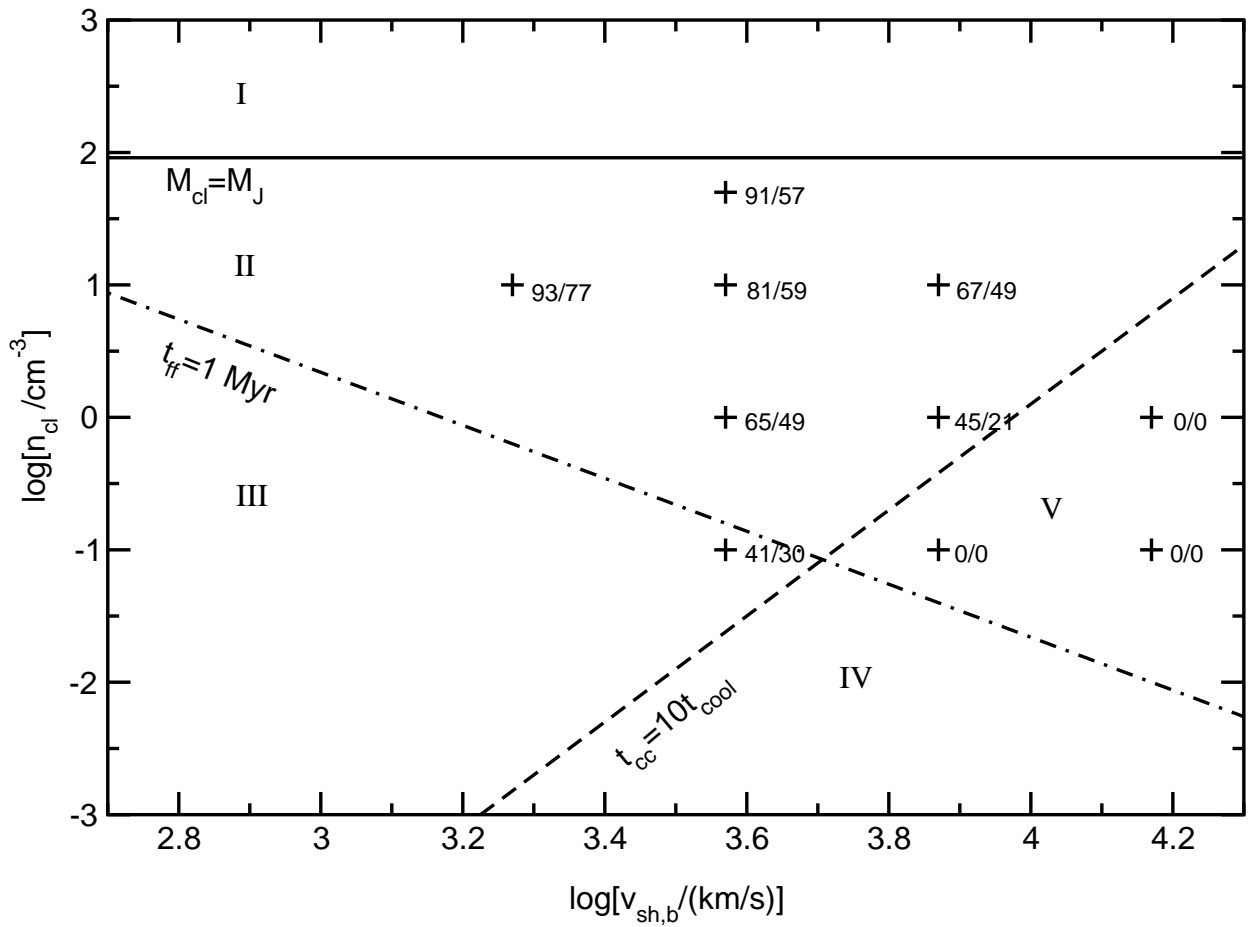}
\caption{General cloud number density ($n_cl$)-shock velocity ($v_{sh,b}$) 
parameter space considered.  The solid line demarks the 
density above which the cloud is initially gravitationally unstable 
(region I).  The dashed line divides the cooling dominated regions 
(II \& III) on the left from the non-cooling regions (IV \& V) on the 
right.  The dot-dashed line is an estimate of the star-formation 
cut-off.  Hence clouds in region III will likely form far fewer stars 
than those found in region II.  The particular parameter pairs explored 
with the equilibrium cooling curve model are indicated 
with crosses.  The two numbers next to each cross give the amount of 
cloud gas that ends the simulation below $T=1000$ and 100 K, respectively, 
as a percent of the initial cloud mass.
}
\label{fig:equil}
\end{figure}

\begin{figure}
%Fig 5 {paramchem_bw.eps}
\epsscale{1.0}
\plotone{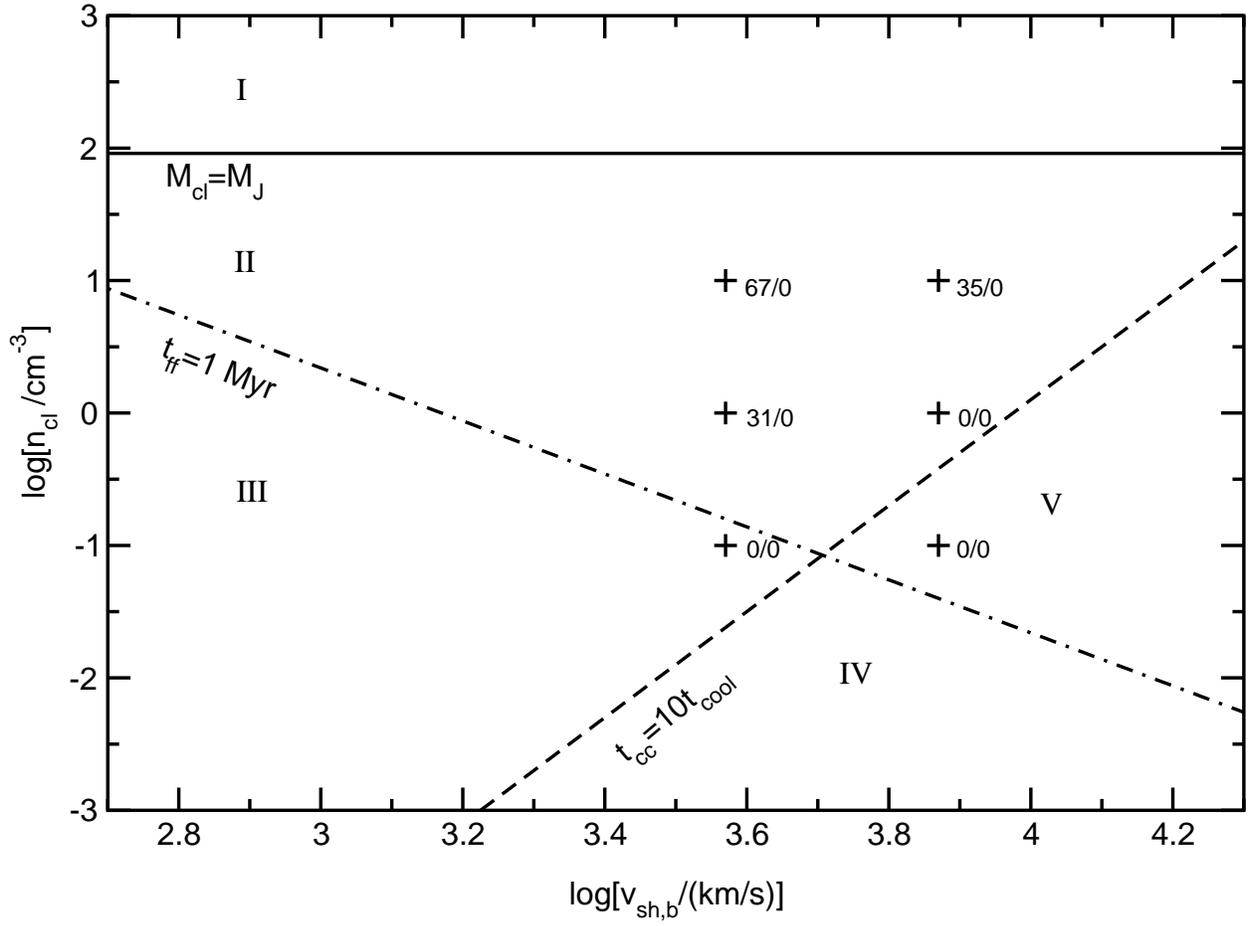}
\caption{Same as Figure \ref{fig:equil} except applies to 
the non-equilibrium 
chemistry model.
}
\label{fig:chem}
\end{figure}

We have performed two-dimensional simulations of the evolution of
radiatively cooling clouds subjected to strong shocks, as might arise from
galactic jets.  The
results of the models are summarized in Figures~\ref{fig:equil} and
\ref{fig:chem}, which show the locations of the models in cloud
density-shock velocity parameter space.
Figure~\ref{fig:equil} presents the results
for models which include equilibrium cooling with solar abundances,
while Figure~\ref{fig:chem} shows the results for
primordial clouds, which include non-equilibrium chemistry
in which cooling at low temperatures is dominated by H$_2$
emission.  The numbers by each model indicate the percentages of the
original gas that cools to below either 1000~K or 100~K.
As can be seen from the figures, large fractions ($\gtrsim 1/3$) of the
clouds are able to cool below 1000~K.  Most of this cold gas ends up in very 
dense filaments, which have long dynamical lifetimes.  The final 
densities of these filaments are high enough in many cases that 
they are gravitationally unstable.  The subsequent gravitational 
collapse of these regions will likely result in a burst of star 
formation.  Our numerical results are consistent with previous 
work \citep[e.g.][]{Mellema02} and have extended it by 
including new physics (chemistry and self-gravity) and 
exploring more of the parameter space of interest.

\begin{figure}
%Fig 1 {Jet-2D-51_den.eps}
\epsscale{0.5}
\plotone{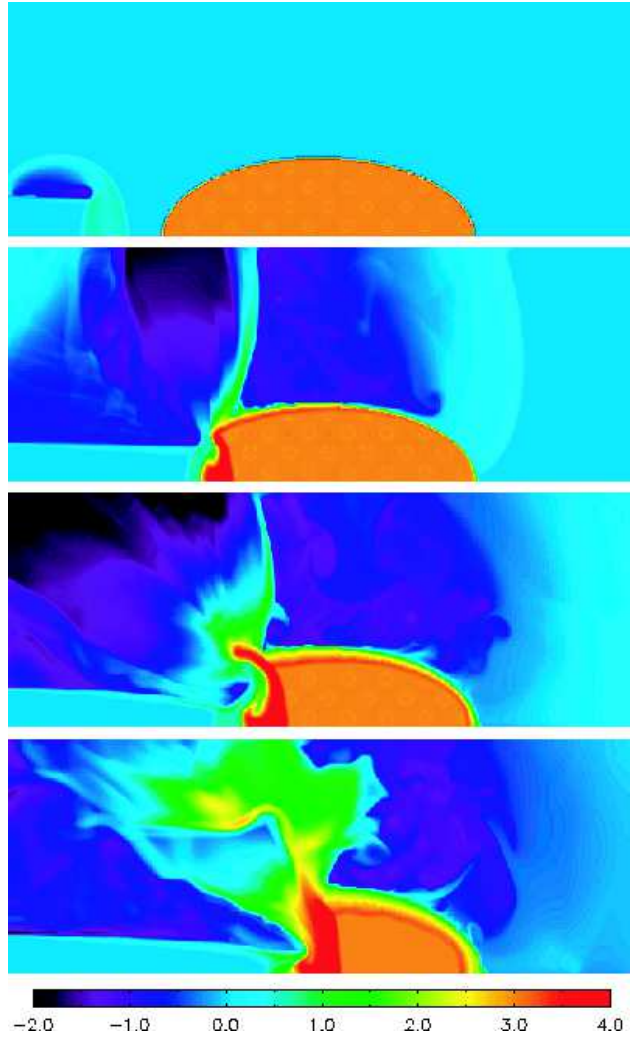}
\caption{Contour plot of the logarithm of gas density for a 
simulation of a jet 
colliding with the long axis of an elliptical cloud 
at simulation times $t=0.1$, 0.8, 1.4, and 2.1 Myr.  
The unit for the density scale 
in this plot is $1.7 \times 10^{-28}$ g cm$^{-3}$, 
corresponding to an initial cloud mass of 
$\approx 10^9 M_\odot$.
}
\label{fig:jetcloud}
\end{figure}

\begin{figure}
%Fig 2 {Minkowski_x2.eps}
\plotone{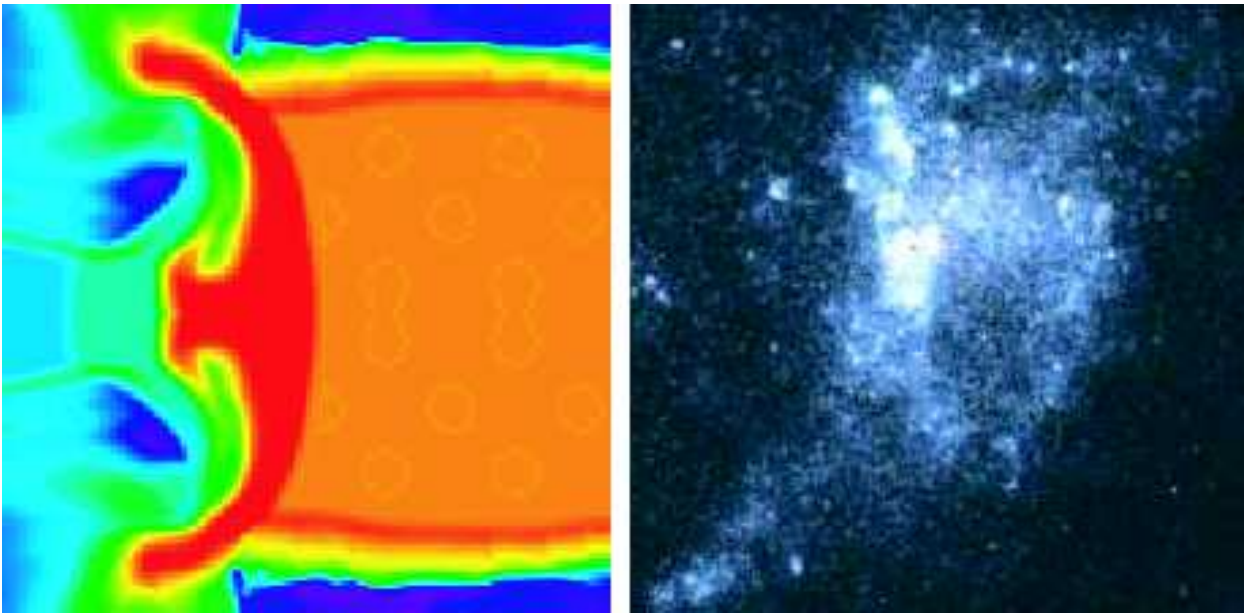}
\caption{Comparison of an intermediate density distribution plot from the 
numerical simulation in Figure \ref{fig:jetcloud} with a similarly scaled 
observation of Minkowski's Object.  There are clear similarities between 
the distribution of the post-shock gas within the simulated cloud 
(colored red) and 
the regions of active star formation within Minkowski's Object.
}
\label{fig:MO}
\end{figure}

We applied our results to the specific example of Minkowski's Object. A
number of important conclusions can be drawn. First, its peculiar
morphology - bright star forming region orthogonal to the jet, and
fainter filamentary features downstream from there - can be easily
reproduced by our simulations (Fig \ref{fig:jetcloud}).  
Second, the modest amount star
formation required - $0.3 M_\odot$ yr$^{-1}$ for the entire object, is
also easily achieved for the plausible parameter space explored by our
simulations.  Third, and most interestingly, we conclude that the star
formation in Minkowski's Object could be induced by a moderate velocity 
jet ($9 \times 10^4$ km s$^{-1}$) interacting with a collection of 
slightly overdense ($\sim 10$ cm$^{-3}$), warm ($10^4$ K) clouds, 
i.e. it is NOT necessary to assume that this was an accidental
collision between a jet and a preexisting gas-rich galaxy. This also
suggests that the neutral hydrogen associated with Minkowski's Object
($3 \times 10^8 M_\odot$; W. van Breugel \& J. van Gorkom 2003, private 
communication) may 
have cooled from the warm gas phase as a result of the radiative cooling
triggered by the radio jet.  An update on the observations of Minkowski's
Object will be presented in a forthcoming paper 
(S. D. Croft et al., in preparation).

Our results show that jets can drive radiative shocks in overdense
regions in the intergalactic medium, resulting in star formation far
away from the galaxies in which the jets originate.  This could be an 
important factor in understanding the feedback of AGN 
on their environment.  Our simulations may
also be applied to radiative shocks and resulting star formation 
triggered by other mechanisms, such as in the wake of supernova blast
waves \citep{Preibisch02}, collisions of interstellar or intergalactic
clouds \citep{Smith80}, and ram pressure on gas-rich galaxies that fall
into galaxy cluster atmospheres \citep[e.g.][]{Gavazzi01}.

Our models can be used to infer the importance of shock-induced star formation
in other regimes by simply 
adjusting the criteria used to develop Figure \ref{fig:equil}.  Our
models have confirmed the reality of the division between regions II and V 
(cooling and non-cooling) derived using approximate analytical
methods, while the other region boundaries are set by
physical limits for the clouds.  In future work, 
we shall extend our numerical simulations
into regimes appropriate for other possible shock-cloud interaction scenarios.

\begin{acknowledgements}
This work was performed under the auspices of the U.S. Department of
Energy by University of California, Lawrence Livermore National
Laboratory under Contract W-7405-Eng-48.  W.v.B.\ also acknowledges
NASA grants GO~9779 and GO3-4150X in support of high-redshift radio galaxy
research with HST and Chandra.
\end{acknowledgements}

%%%%%%%%%%%%%%%%%%%% TABLES %%%%%%%%%%%%%%%%%%%%%

\clearpage
\begin{deluxetable}{cccccccl}
\tablewidth{0pt}
\tablecaption{Equilibrium Cooling Curve Models\label{tab:equil}}
\tablehead{
\colhead{} &
\colhead{$n_{cl}$} &
\colhead{$v_{sh,b}$} &
\colhead{} &
\colhead{$t_{sp}$} &
\colhead{$t_{cc}$} &
\colhead{$t_{cool}$} &
\colhead{} \\
\colhead{Model} &
\colhead{(cm$^{-3}$)} &
\colhead{(10$^3$ km s$^{-1}$)} &
\colhead{$\mathcal{M}$} &
\colhead{(yr)} &
\colhead{(yr)} &
\colhead{(yr)} &
\colhead{Notes\tablenotemark{a}} 
}
\startdata
E1a & 50 & 3.7 & 10 & $5.3 \times 10^4$ & $8.5 \times 10^5$ & $3.9 \times 10^1$ & \\
E1b & 50 & 3.7 & 10 & $5.3 \times 10^4$ & $8.5 \times 10^5$ & $3.9 \times 10^1$ & self gravity \\
E2 & 10 & 1.9 & 5 & $1.1 \times 10^5 $ & $1.7 \times 10^6$ & $2.4 \times 10^1$ & \\
E3 & 10 & 3.7 & 10 & $5.3 \times 10^4$ & $8.5 \times 10^5$ & $1.9 \times 10^2$ & (C1a) \\
E4 & 10 & 7.4 & 20 & $2.7 \times 10^4$ & $4.3 \times 10^5$ & $1.6 \times 10^3$ & (C2) \\
E5 & 1 & 3.7 & 10 & $5.3 \times 10^4$ & $8.5 \times 10^5$ & $1.9 \times 10^3$ & (C3) \\
E6a & 1 & 7.4 & 20 & $2.7 \times 10^4$ & $4.3 \times 10^5$ & $1.6 \times 10^4$ & (C4) \\
E6b & 1 & 7.4 & 20 & $2.7 \times 10^4$ & $4.3 \times 10^5$ & $1.6 \times 10^4$ & multi-cloud \\
E7 & 1 & 14.9 & 40 & $1.3 \times 10^4$ & $2.1 \times 10^5$ & $1.2 \times 10^5$ & \\
E8 & 0.1 & 3.7 & 10 & $5.3 \times 10^4$ & $8.5 \times 10^5$ & $1.9 \times 10^4$ & (C5) \\
E9 & 0.1 & 7.4 & 20 & $2.7 \times 10^4$ & $4.3 \times 10^5$ & $1.6 \times 10^5$ & (C6) \\
E10 & 0.1 & 14.9 & 40 & $1.3 \times 10^4$ & $2.1 \times 10^5$ & $1.2 \times 10^6$ & \\
\enddata
\tablenotetext{a}{The model name in parentheses corresponds to the 
non-equilibrium chemistry model in Table \ref{tab:chem} with the same 
choices of parameters, $n_{cl}$ and $v_{sh,b}$.}
\end{deluxetable}

\clearpage
\begin{deluxetable}{ccccccl}
\tablewidth{0pt}
\tablecaption{Non-equilibrium Chemistry Models\label{tab:chem}}
\tablehead{
\colhead{} &
\colhead{$n_{cl}$} &
\colhead{$v_{sh,b}$} &
\colhead{} &
\colhead{$t_{sp}$} &
\colhead{$t_{cc}$} &
\colhead{} \\
\colhead{Model} &
\colhead{(cm$^{-3}$)} &
\colhead{(km s$^{-1}$)} &
\colhead{$\mathcal{M}$} &
\colhead{(yr)} &
\colhead{(yr)} &
\colhead{Notes\tablenotemark{a}} 
}
\startdata
C1a & 10 & 3.7 & 10 & $5.3 \times 10^4$ & $8.5 \times 10^5$ & (E3) \\
C1b & 10 & 3.7 & 10 & $5.3 \times 10^4$ & $8.5 \times 10^5$ & No photo-ionization or photo-dissociation \\
C1c & 10 & 3.7 & 10 & $5.3 \times 10^4$ & $8.5 \times 10^5$ & No optical depth calculations \\
C1d & 10 & 3.7 & 10 & $5.3 \times 10^4$ & $8.5 \times 10^5$ & Photoionizing field 10 times stronger \\
C1e & 10 & 3.7 & 10 & $5.3 \times 10^4$ & $8.5 \times 10^5$ & Photodissociation rate 10 times higher \\
C2 & 10 & 7.4 & 20 & $2.7 \times 10^4$ & $4.3 \times 10^5$ & (E4) \\
C3 & 1 & 3.7 & 10 & $5.3 \times 10^4$ & $8.5 \times 10^5$ & (E5) \\
C4 & 1 & 7.4 & 20 & $2.7 \times 10^4$ & $4.3 \times 10^5$ & (E6a) \\
C5 & 0.1 & 3.7 & 10 & $5.3 \times 10^4$ & $8.5 \times 10^5$ & (E8) \\
C6 & 0.1 & 7.4 & 20 & $2.7 \times 10^4$ & $4.3 \times 10^5$ & (E9) \\
\enddata
\tablenotetext{a}{The model name in parentheses corresponds to the 
equilibrium cooling-curve  model in Table \ref{tab:equil} with the same 
choices of parameters, $n_{cl}$ and $v_{sh,b}$.}
\end{deluxetable}

\clearpage
\begin{deluxetable}{ccc}
\tablewidth{0pt}
\tablecaption{Equilibrium Cooling Curve Results \label{tab:cooleq}}
\tablehead{
\colhead{} &
\multicolumn{2}{c}{Gas with} \\
\cline{2-3} \\
\colhead{} &
\colhead{$T_f<1000$ K} &
\colhead{$T_f<100$ K} \\
\colhead{Model} &
\colhead{(\% of $M_{cl}$)} &
\colhead{(\% of $M_{cl}$)} 
}
\startdata
E1a & 91 & 57 \\
E1b & 92 & 69 \\
E2 & 93 & 77 \\
E3 & 81 & 59 \\
E4 & 67 & 49 \\
E5 & 65 & 49 \\
E6a & 45 & 21 \\
E6b\tablenotemark{a} & 46,26,21 & 9,16,17 \\
E7 & 0 & 0 \\
E8 & 41 & 30 \\
E9 & 0 & 0 \\
E10 & 0 & 0 \\
\enddata
\tablenotetext{a}{Results given separately for each cloud, moving 
from left to right in Figure \ref{fig:multi}.}
\end{deluxetable}

\clearpage
\begin{deluxetable}{cccccc}
\tablewidth{0pt}
\tablecaption{Non-Equilibrium Chemistry Results \label{tab:coolch}}
\tablehead{
\colhead{} &
\multicolumn{2}{c}{Gas with} &
\colhead{} &
\multicolumn{2}{c}{H$_2$ Mass Fraction} \\
\cline{2-3} \cline{5-6} \\
\colhead{} &
\colhead{$T_f<1000$ K} &
\colhead{$T_f<100$ K} &
\colhead{} &
\colhead{Peak} &
\colhead{Total} \\
\colhead{Model} &
\colhead{(\% of $M_{cl}$)} &
\colhead{(\% of $M_{cl}$)} &
\colhead{} &
\colhead{($\rho_{H_2}/\rho$)} &
\colhead{($M_{H_2}/M_{cl}$)} 
}
\startdata
C1a & 67 & 0 & & $1.3 \times 10^{-2}$ & $1.3 \times 10^{-5}$ \\
C1b & 72 & 0 & & $4.0 \times 10^{-2}$ & $3.9 \times 10^{-5}$ \\
C1c & 13 & 0 & & $6.3 \times 10^{-4}$ & $1.0 \times 10^{-6}$ \\
C1d & 69 & 0 & & $3.2 \times 10^{-2}$ & $2.5 \times 10^{-5}$ \\
C1e & 53 & 0 & & $1.0 \times 10^{-2}$ & $3.9 \times 10^{-6}$ \\
C2  & 35 & 0 & & $1.0 \times 10^{-2}$ & $8.3 \times 10^{-6}$ \\
C3  & 31 & 0 & & $5.0 \times 10^{-3}$ & $6.9 \times 10^{-7}$ \\
C4  & 0  & 0 & & $4.0 \times 10^{-4}$ & $4.6 \times 10^{-7}$ \\
C5  & 0  & 0 & & $1.0 \times 10^{-4}$ & $7.6 \times 10^{-8}$ \\
C6  & 0  & 0 & & $1.3 \times 10^{-4}$ & $3.9 \times 10^{-8}$ \\
\enddata
\end{deluxetable}

%%%%%%%%%%%%%%%%%%%%%% FIGURES %%%%%%%%%%%%%%%%%

\clearpage

\clearpage
\epsscale{1.0}

\clearpage

\clearpage

\clearpage

\clearpage

\clearpage


\clearpage
\begin{thebibliography}{}

\bibitem[Anninos \& Fragile (2003)]{AF03} Anninos, P., \& 
Fragile, P. C. 2003, \apjs, 144, 243

\bibitem[Anninos, Fragile \& Murray (2003)]{AFM03} Anninos, P., Fragile,
P. C., \& Murray, S. D. 2003, \apjs, 147, 177

\bibitem[Anninos et al. (1997)]{Anninos97} 
Anninos, P., Zhang, Y., Abel, T., \& Norman, M. L. 1997, 
New Astronomy, 2, 209

\bibitem[Bechtold et al. (1987)]{Bechtold87} Bechtold, J., Weymann, R. J.,
Lin, Z., \& Malkan, M. A. 1987, \apj, 315, 180

\bibitem[Begelman \& Cioffi (1989)]{Begelman89}
Begelman, M. C. \& Cioffi, D. F. 1989, \apj, 345, L21

\bibitem[Bicknell (1984)]{bic84}
Bicknell, G. V. 1984, \apj, 286, 68

\bibitem[Bicknell et al. (2000)]{Bicknell00}
Bicknell, G. V., Sutherland, R. S., van Breugel, W. J. M., Dopita, M. A., 
Dey, A., \& Miley, G. K. 2000, \apj, 540, 678

\bibitem[Blanco et al. (1975)]{Blanco75}
Blanco, V. M., Graham, H. A., Lasker, B. M., \& Osmer, P. S. 1975, \apj, 
198, L63

\bibitem[Chambers, Miley, \& van Breugel (1987)]{Chambers87}
Chambers, K. C., Miley, G. K., \& van Breugel, W. 1987, Nature, 329, 604

\bibitem[Chandrasekhar (1961)]{Chandra61}
Chandrasekhar, S. 1961, Hydrodynamic and Hydromagnetic Stability 
(New York: Dover)

\bibitem[Dalgarno \& McCray (1972)]{Dalgarno72} Dalgarno, A., \& 
McCray, R. A. 1972, \araa, 10, 375

\bibitem[Dey et al. (1997)]{Dey97}
Dey, A., van Breugel, W., Vacca, W. D., \& Antonucci, R. 1997, \apj, 490, 698

\bibitem[De Young (1989)]{DeYoung89}
De Young, D. S. 1989, \apj, 342, L59

\bibitem[Dong, Lin, \& Murray (2003)]{DLM03} Dong, S., Lin, D. N. C.,
\& Murray, S. D. 2003, \apj, in press

\bibitem[Fanaroff \& Riley (1974)]{Fanaroff74}
Fanaroff, B. L. \& Riley, J. M. 1974, \mnras, 167, 31P

\bibitem[Ferland, Fabian, \& Johnstone (2002)]{Ferland02}
Ferland, G. J., Fabian, A. C., \& Johnstone, R. M. 2002, \mnras, 333, 876

\bibitem[Field (1965)]{Field65} Field, G. B. 1965, \apj, 142, 531

\bibitem[Gavazzi et al. (2001)]{Gavazzi01}
Gavazzi, G., Boselli, A., Mayer, L., Iglesias-Paramo, J., V\'ilchez, J. M., 
\& Carrasco, L. 2001, \apj, 563, L23

\bibitem[Gregori et al. (1999)]{Gregori99} Gregori, G., Miniati, F., 
Ryu, D., \& Jones, T. W. 1999, \apj, 527, L113

\bibitem[Hartquist \& Dyson (1988)]{Hartquist88} Hartquist, T. W. 
\& Dyson, J. E. 1988, \apss, 144, 615

\bibitem[Hollenbach, Werner, \& Salpeter (1971)]{Hollenbach71}
Hollenbach, D. J., Werner, M. W., \& Salpeter, E. E. 1971, \apj, 163, 165

\bibitem[Kahn (1976)]{Kahn76} Kahn, F. D. 1976, \aap, 50, 145

\bibitem[Klein, McKee, \& Colella (1994)]{Klein94} Klein, R. I., 
McKee, C. F., \& Colella, P. 1994, \apj, 420, 213

\bibitem[Laing \& Bridle (2002)]{Laing02}
Laing, R. A. \& Bridle, A. H. 2002, \mnras, 336, 1161

\bibitem[Lepp \& Shull (1983)]{LS83} Lepp, S.
\& Shull, J. M. 1983, \apj, 270, 578

\bibitem[Lin \& Murray (2000)]{LinMurray00} Lin, D. N. C. \&
Murray, S. D. 2000, \apj, 540, 170

\bibitem[McCarthy et al. (1987)]{McCarthy87}
McCarthy, P. J., van Breugel, W. J. M., Spinrad, H., \& Djorgovski, S. 
1987, \apj, 321, L29

\bibitem[McNamara (2002)]{McNamara02}
McNamara, B. R. 2002, New Astronomy Review, 46, 141

\bibitem[Mellema, Kurk, \& R\"ottgering (2002)]{Mellema02} 
Mellema, G., Kurk, J. D., R\"ottgering, H. J. A. 2002, \aap, 395, L13

\bibitem[Mould et al. (2000)]{Mould00}
Mould, J. R. et al. 2000, \apj, 536, 266

\bibitem[Murray \& Lin (1989)]{MurrayLin89} Murray, S. D. \& Lin,
D. N. C. 1989, \apj, 339, 933 (see also Erratum, 1989, \apj, 344, 1052)

\bibitem[Osterbrock (1989)]{Osterbrock89} Osterbrock, D. E. 1989,
Astrophysics of Gaseous Nebulae and Active Galactic Nuclei (Mill Valley:
University Science Books)

\bibitem[Poludnenko, Frank, \& Blackman (2002)]{Poludnenko02}
Poludnenko, A. Y., Frank, A., \& Blackman, E. G. 2002, \apj, 576, 832

\bibitem[Preibisch et al. (2002)]{Preibisch02}
Preibisch, T., Brown, A. G. A., Bridges, T., Guenther, E., \& 
Zinnecker, H. 2002, \apj, 124, 404

\bibitem[Rees (1989)]{Rees89}
Rees, M. J. 1989, \mnras, 239, 1P

\bibitem[Rejkuba et al. (2002)]{rej02}
Rejkuba, M., Minniti, D., Courbin, F., \& Silva, D. R. 2002, \apj, 564, 688

\bibitem[Reuland et al. (2003)]{Reuland03}
Reuland, M. et al. 2003, \apj, 592, 755

\bibitem[Sgro (1975)]{Sgro75} Sgro, A. G. 1975, \apj, 197, 621

\bibitem[Smith (1980)]{Smith80} Smith, J. 1980, \apj, 238, 842

\bibitem[Spaans \& Neufeld (1997)]{Spaans97} Spaans, M. \& Neufeld, D. A.
1997, \apj, 484, 785

\bibitem[van Breugel et al. (1985)]{vB85}
van Breugel, W., Filippenko, A. V., Heckman, T., \& Miley, G. 1985, 
\apj, 293, 83

\bibitem[van Breugel et al. (1999)]{vB99}
van Breugel, W., Stanford, A., Dey, A., Miley, G., Stern, D., Spinrad, H., 
Graham, J., \& McCarthy, P. 1999, in The Most Distant Radio Galaxies, 
ed. H. J. A. Röttgering, P. N. Best, \& M. D. Lehnert 
(Amsterdam: Royal Netherlands Academy of Arts and Sciences), 49

\end{thebibliography}
\end{document}